\documentclass[a4paper, 11pt]{article}
\usepackage{jheppub}
\usepackage[T1]{fontenc} 
\usepackage{graphicx}
\usepackage{tikz}
\usepackage{amsmath,amssymb,multirow,xspace,slashed,array,booktabs}
\usepackage[colorlinks=true,urlcolor=blue,anchorcolor=blue,citecolor=blue,filecolor=blue,linkcolor=blue,menucolor=blue,pagecolor=blue]{hyperref}
\usepackage{natbib}
\setcitestyle{square, comma, numbers,sort&compress, super}
\usepackage{booktabs}
\usepackage{blindtext, rotating}
\usepackage{appendix}
\usepackage{afterpage}
\usepackage{changepage}
\usepackage{enumitem}
\usepackage{marvosym}
\usepackage{bbold}
\usepackage{slashed}
\usepackage{physics}
\usepackage{verbatim}
\usepackage{indentfirst}
\usepackage{bm}
\usepackage{soul} 
\usepackage[normalem]{ulem}
\usepackage{pifont}
\usepackage{cancel}
\usepackage{float}
\usepackage{caption}
\usepackage{subcaption}
\usepackage{cleveref}
\crefname{figure}{Fig.}{Figs.}
\Crefname{figure}{Fig.}{Figs.}

\crefrangeformat{figure}{Figs.~#3#1#4--#5#2#6}
\Crefrangeformat{figure}{Figs.~#3#1#4--#5#2#6}

\title{Filling the Gap: Hunting for Vector Bosons at the MUonE Experiment with Displaced Decay Signature}

\author[a]{Duncan Rocha}
\author[b]{Isaac R. Wang}

\affiliation[a]{Enrico Fermi Institute, University of Chicago, Chicago, IL USA}
\affiliation[b]{Theoretical Physics Division, Fermi National Accelerator Laboratory, Batavia, IL, USA}

\emailAdd{drocha@uchicago.edu}
\emailAdd{isaacw@fnal.gov}
\preprint{FERMILAB-PUB-25-0795-T}

\abstract{
The upcoming MUonE experiment aims to precisely measure the running of the fine structure constant via elastic muon-electron scattering, to shed light on the current tension in the muon's anomalous magnetic moment.
In addition to its primary function as a precision experiment, MUonE also offers a unique testing ground to probe long-lived vector bosons.
Such vector bosons can be produced via $\mu e \to \mu e V$ or $\mu N \to \mu N V$ scattering and decay into an electron/positron pair a few centimeters away from the interaction point.
With its high-resolution tracking system and unique geometric design, MUonE is well-suited to reconstruct displaced vertices close to the target, allowing it to probe parameter space previously unattainable at colliders and longer-baseline beam dump experiments.
We present a comprehensive study of the discovery potential of BSM vector boson mediators at the MUonE experiment. 
We show that MUonE can fill the long-standing gap in the parameter space of vector boson mediators with masses up to around 100 MeV. 
}
\begin{document}
\maketitle
\flushbottom

\section{Introduction}

Since the development of the Standard Model (SM), numerous Beyond-Standard-Model (BSM) force carriers have been proposed, including the dark photon~\cite{Okun:1982xi,Galison:1983pa, Holdom:1985ag, Boehm:2003hm, Pospelov:2008zw}, gauged lepton number~\cite{Foot:1990mn,He:1990pn,He:1991qd}, gauged $B-L$~\cite{Davidson:1978pm,Marshak:1979fm,Mohapatra:1980qe}, extra scalars~\cite{Anderson:1991zb,Pietroni:1992in,Espinosa:1993bs,McDonald:1993ey,Choi:1993cv,Ham:2004cf,Patt:2006fw,OConnell:2006rsp}, along with other scenarios.
Beyond serving as simple extensions to the SM, hidden forces can address several open questions in modern physics.
In particular, a hidden force mediator coupled to a dark sector can naturally account for the observed abundance of dark matter through the thermal freeze-out mechanism~\cite{Essig:2013lka,Alexander:2016aln, Battaglieri:2017aum, Beacham:2019nyx, Alimena:2019zri,Alimena:2021mdu,Bernal:2017kxu,Curtin:2018mvb}.
It remains undetermined whether any hidden forces exist in nature, although many experiments have been conducted to probe such models.
If produced in an experiment, hidden forces could manifest in one of three possible regimes. In one regime, the mediators decay back to SM particles promptly, requiring a resonance search. If there are dark sector particles less massive than the mediator, the mediator may instead decay into the dark sector invisibly, requiring a search for missing energy. Most relevant for this work is the regime where the mediator decays visibly at a location displaced from the interaction point, requiring precise vertex reconstruction to detect.
See Ref.~\cite{Ilten:2018crw,Bauer:2018onh,Graham:2021ggy,Krnjaic:2025noj} for systematic studies.

Interestingly, an experimental ``gap'' exists in parameter space where the hidden force coupling is too small for the mediator to be produced at colliders, but it also is not long-lived enough to provide a signal in existing beam dump experiments.
For characteristic beam energies of $\mathcal O(100)$\,GeV, a hidden boson in the ``gap'' would produce a displaced vertex a few centimeters away from the interaction point, and thus requires a compact detector geometry to be identified.

The MUonE experiment~\cite{MUonE:2016hru,MUonE:2019qlm} was recently shown to be sensitive to the dark photon model in the ``gap''~\cite{Galon:2022xcl}\footnote{For other proposals of searching for gauge bosons at MUonE, see Ref.~\cite{Dev:2020drf,Masiero:2020vxk,Asai:2021wzx,Asai:2023mzl}.}.
MUonE was originally proposed to measure elastic $\mu e$ scattering and extract the hadronic vacuum polarization contribution to the running of $\alpha$.
To make this measurement, MUonE has developed a tracking system capable of reconstructing charged tracks with precise angular resolution~\cite{MUonE:2016hru,MUonE:2019qlm}.
This state-of-the-art tracking system also enables MUonE to detect displaced vertices, making it a promising platform for a long-lived particle search.
Ref.~\cite{Galon:2022xcl} studied dark photon production through $\mu e \to \mu e A’$ with $A’ \to e^+e^-$, considering both a 4-track search including all final state leptons and a 3-track search where the primary electron need not be observed.
Using a simulation-based analysis, they illustrated that MUonE can probe regions in the ``gap'' for dark photon masses up to 70 MeV.

Importantly, Ref.~\cite{Galon:2022xcl} showed that the SM background can still be identified and rejected even if the primary electron is not detected, which opens a new possibility of considering the elastic production channel where a muon scatters off the nucleus and radiates a dark vector boson.
The much larger center-of-mass (CoM) energy in $\mu N$ scattering over $\mu e$ can significantly enhance the maximum vector boson mass observable at MUonE.
This was previously applied to show that MUonE is sensitive to inelastic dark matter in the ``gap'' up to 700 MeV~\cite{Krnjaic:2024ols}.

At the time of writing this paper, MUonE finished its test run and is currently analyzing the data collected.
The next stage is an official run of the experiment, projected to begin in early 2027~\cite{Umberto:2021private}.
This recent progress calls for a detailed assessment of the sensitivity of MUonE to hidden force models.
In this work, we consider three well-motivated vector boson models and generalize the displaced vertex search at MUonE to include both $\mu N$ and $\mu e$ scattering production channels.
We conduct a simulation-based analysis and implement a search strategy following Ref.~\cite{Galon:2022xcl,Krnjaic:2024ols}, aimed at making reliable projections.
Our main result is a characterization of the sensitivity of MUonE to the existence of a dark photon, $U(1)_{B-L}$ gauge boson, and $U(1)_{L_e - L_\mu}$ gauge boson.

This paper is organized as follows.
In Sec.~\ref{sec:model}, we provide a short review of the three models of vector boson mediators studied.
In Sec.~\ref{sec:signal}, we discuss how these mediators can be produced and detected at MUonE, with description of event generation and background simulation.
In Sec.~\ref{sec:results}, we present the projected sensitivity curve for the three models.
We conclude this paper in Sec.~\ref{sec:conclusion}.

\section{Simple Models of Vector Boson Mediators}
\label{sec:model}
We consider hidden force models that include a new massive $U(1)$ vector boson $V_\mu$ which couples to some (or all) of the SM fermions.
The interaction term in the low energy limit is
\begin{align}
    \mathcal{L} \supset g_V \sum_{f\in \rm SM} c_f \bar{f} \gamma^\mu V_\mu f\,,
\end{align}
where the sum is over all SM fermions. Here, $g_V c_f$ is the coupling between the massive vector boson and SM fermion $f$.
Additional dark fermions or scalars may be present and charged under the new hidden force, thus forming a ``dark sector.''
We assume that any such fields are more massive than the mediator, so that the force carrier $V$ decays only into SM products.

Notably, the list of viable $U(1)$ gauge symmetries that can be consistently added to the SM is severely restricted by anomaly cancellation. The models we consider can all be consistently added without requiring additional fields to cancel gauge anomalies, making them well-motivated and flexible extensions of the SM.

\begin{itemize}
    \item \textbf{Dark photon}: this scenario posits a $U(1)_D$ under which no SM fields are charged. However, heavy particles charged under both $U(1)_D$ and $U(1)_{EM}$ can induce a small gauge-invariant kinetic mixing term in the IR. Therefore, the mediator $A_\mu'$ of the $U(1)_D$ inherits the same structure of charge assignments as the SM photon~\cite{Holdom:1986eq}. It couples to all charged particles in SM with strength $\epsilon e Q$, where $\epsilon$ is the mixing parameter, $e$ is the QED coupling, and $Q$ is the electric charge of the charged particle.
    \item \textbf{Gauged \boldmath $U(1)_{B-L}$}: this scenario gauges the existing global $U(1)_{B-L}$ symmetry in the SM~\cite{Davidson:1978pm,Marshak:1979fm,Mohapatra:1980qe}. This avoids a gauge anomaly if one assumes right-handed neutrinos exist, and we take them to have heavy Majorana masses. The $Z'_{\mu, B-L}$ gauge boson couples to all quarks (leptons) with coupling $g_{B-L}/3$ ($g_{B-L}$).
    \item \textbf{Gauged \boldmath $U(1)_{L_e - L_\mu}$}: this scenario gauges a difference of $L_e$ and $L_\mu$~\cite{Foot:1990mn}.
          The $Z_{e\mu}'$ gauge boson couples to the muon (muon neutrino) and electron (electron neutrino) with coupling $\pm g_{e\mu}$, respectively.
          In principle, any of the three combinations can be chosen, but $U(1)_{L_{e} - L_{\mu}}$ is the model for which MUonE is especially sensitive.
\end{itemize}
Their properties are summarized in Table.~\ref{tab:model}.
\begin{table}[t!]
    \centering
    \hspace*{-0.65cm}
    \begin{tabular}{|c|c|c|c|c|c|}
        \hline
                             & $g_V$        & $c_u$ & $c_d$  & $c_\nu$                                                 & $c_\ell$                              \\
        \hline
        Dark photon          & $\epsilon e$ & $2/3$ & $-1/3$ & $0$                                                     & $-1$                                  \\
        \hline
        Gauged $L_e - L_\mu$ & $g_{e\mu}$   & 0     & 0      & $c_{\nu_e} = 1$, $c_{\nu_\mu} = -1$, $c_{\nu_\tau} = 0$ & $c_e = 1$, $c_\mu = -1$, $c_\tau = 0$ \\
        \hline
        Gauged $B - L$       & $g_{B-L}$    & $1/3$ & $1/3$  & $-1$                                                    & $-1$                                  \\
        \hline
    \end{tabular}
    \caption{The vector boson mediator considered in this work. Columns show the notation of the coupling $g_V$ and the charge of each SM fermions, respectively.}
    \label{tab:model}
\end{table}

\begin{figure}[t!]
    \centering
    \includegraphics[width=0.9\linewidth]{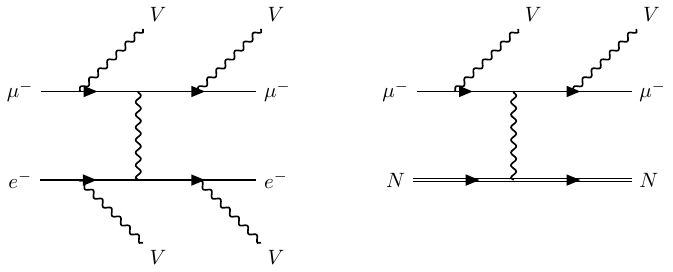}
    \caption{Feynman diagrams of vector boson production at MUonE.}
    \label{fig:feynman}
\end{figure}

Fig.~\ref{fig:feynman} shows the Feynman diagrams for vector boson production at MUonE.
All mediators can be produced by ``dark bremsstrahlung'' in $\mu e$ or $\mu N$ scattering.
For $\mu N$ scattering, bremsstrahlung off of the nucleus itself is severely suppressed, and thus for a given vector boson mass the cross section can be straightforwardly rescaled between models, i.e.
\begin{align}
    \label{eq:xsec}
    \frac{\sigma (\mu N \to \mu N V)}{\sigma (\mu N \to \mu N A')}  = \frac{(g_V c_\mu)^2}{(\epsilon e)^2}\,,
\end{align}
where $V$ stands for a general vector mediator.
$\epsilon e$ can be any reference value in principle, and is chosen to be $\epsilon = 10^{-3}$ during the production simulation.
On the contrary, vector bosons from $\mu e$ scattering can be emitted from both the muon and electron legs.
The interference behavior changes depending on the relative sign of the $Vee$ and $V\mu\mu$ vertices, an effect which we find numerically to be an $\mathcal O(10)\%$ change in cross section. This effect can be seen in Fig.~\ref{fig:productionCrossSections}.

The decay rate of the vector boson $V_\mu$ into a pair of SM fermions in the CoM frame is
\begin{align}
    \label{eq:decay fermion}
    \Gamma_{\bar{f}f} = \frac{N_f (g_V c_f)^2}{12 \pi} m_V \left( 1 + \frac{2m_f^2}{m_V^2} \right) \sqrt{1 - \frac{4m_f^2}{m_V^2}}\,.
\end{align}
In the region excluded by MUonE, the mass of the vector boson is below $2 m_\mu$ and $2 m_\pi$, so the only relevant decays are to electrons, neutrinos, or photons.
For $e$ and $\mu$, $N_f = 1$. For $\nu$, $N_f = 1/2$.
Assuming only one decay channel into a pair of Dirac fermions ($N_f = 1$) with $m_f \ll m_V$,
the decay length in the lab frame is given by
\begin{align}
    d_V = \frac{|\vec p_V|}{m_V \Gamma_V} \simeq 74~\mathrm{mm} \left( \frac{10^{-5}}{g_V c_f} \right)^2 \left( \frac{p_V}{10~\rm GeV} \right) \left( \frac{100~\rm MeV}{m_V} \right)^2\,,
\end{align}
where $\vec p_V$ is the momentum of the vector boson in the lab frame, so that $|\vec p_V|/m_V$ represents the boost factor times velocity.
The decay length of other vector bosons can be computed in a similar way, with the width summed over all possible final states.

\section{Vector Boson Signatures at the \textit{MUonE} Experiment}
\label{sec:signal}
In this section, we introduce the signal of displaced decays of BSM vector bosons at the MUonE experiment,
following Ref.~\cite{Galon:2022xcl,Krnjaic:2024ols}.

\subsection{The Apparatus}

The MUonE experiment aims to measure the hadronic vacuum polarization (HVP) contribution to the running of $\alpha$~\cite{MUonE:2016hru,MUonE:2019qlm}.
To do so, MUonE will follow the strategy put forth by the NA7 experiment~\cite{Amendolia:1984nz,NA7:1986vav}.
This approach uses precise measurements of the $\mu e$ elastic scattering cross section to extract the HVP contribution to $\alpha$, providing an important cross-check with the result from lattice QCD~\cite{Borsanyi:2020mff}.
This quantity is crucial for making a precise theoretical prediction of the muon anomalous magnetic moment~\cite{Aoyama:2020ynm,Boccaletti:2024guq} and investigating the current tension with experimental measurements~\cite{Muong-2:2006rrc,Muong-2:2021ojo,Muong-2:2023cdq}.

\begin{figure}[t!]
    \centering
    \includegraphics[width=0.91\linewidth]{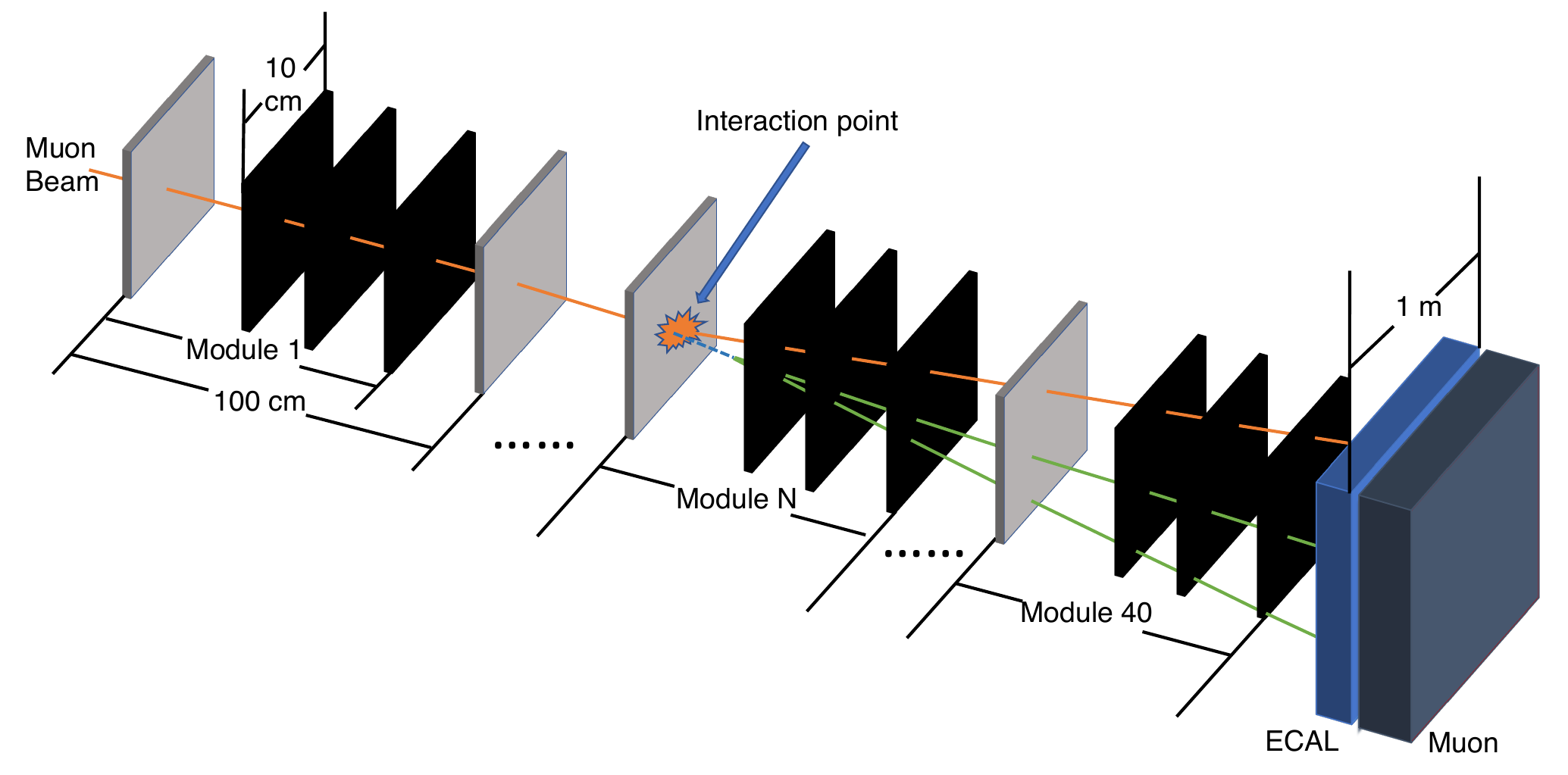}
    \caption{A schematic diagram for the MUonE apparatus with displaced vertex signature, borrowed from Fig.~\cite{Krnjaic:2024ols}. Orange line: incoming muon beam and outgoing scattered muon. Blue dashed line: displaced decaying vector boson. Green lines: decay products of the vector boson. Black squares: tracking stations. Gray squares: Beryllium targets. Blue and dark blue squares: the ECAL and the muon filter.}
    \label{fig:apparatus}
\end{figure}

To achieve this measurement, MUonE delivers a 160 GeV muon beam from the CERN M2 beamline onto 1.5\,cm thick Beryllium (Be) targets.
The targets are put into 40 identical modules aligned with the beamline, each separated by 1\,m.
Each module contains one Be target and three 10\,cm\,$\times$\,10\,cm tracking stations, with the first station located 15\,cm away from the target.
The location of the 2nd and 3rd tracking stations has not been precisely determined yet.
We assume the 3rd one to be located 1\,m away from the target, while the 2nd one stands in the middle between the 1st one and the 3rd one.
More tracking stations may be inserted in the middle to ensure achievement of the proposed angular and vertex reconstruction resolution.
At the very end of the tracking stations, an electromagnetic calorimeter (ECAL) and a muon filter with size 1\,m\,$\times$\,1\,m are placed for the purpose of particle identification (PID).
With 2 years of data taking, the proposed total luminosity is $1.5 \times 10^4~\rm pb^{-1}$ for $\mu e$ scattering, corresponding to a total of $\sim 10^{16}$ muons on target.

The elastic $\mu e \to \mu e$ scattering can be distinguished from other events since the outgoing muon angle $\theta_\mu$ and electron angle $\theta_e$ strictly follow a one-on-one correspondence, known as the elastic curve~\cite{Banerjee:2020tdt}.
To utilize the elastic curve for background rejection,
MUonE employs a CMS-based high-precision tracking system for charged track reconstruction~\cite{MUonE:2019qlm,CERN-LHCC-2017-009,Migliore:2797715}, which achieves an angular resolution within $0.02\,\rm mrad$ of the outgoing particles.
No magnetic field is applied, so charge discrimination and momentum measurement are not possible.
The energy of the outgoing muon is not measured, and the energy resolution of the ECAL has not been finally determined yet.
For this reason, a missing energy search is not possible in MUonE.

Fig.~\ref{fig:apparatus} shows a schematic picture of the MUonE apparatus together with our proposed displaced vertex signature.
The signal event topology is discussed in the next subsection.

\subsection{Search Strategy}
\label{sec:search}

If a vector bosons from a hidden force are produced in MUonE, signal events would include a pair of charged tracks originating from a displaced vertex, corresponding to the decay of the vector boson.
The final state thus contains three or four charged tracks: one coming from the primary interaction vertex, identified as the deflected muon, two reconstructed to be coming from a common displaced vertex and identified as a pair of electrons, and the potentially fourth one coming from the atomic electron that may also be involved in the production mechanism.

\begin{figure}[t!]
    \centering
    \vspace{-0.5cm}
    \includegraphics[width=0.7\linewidth]{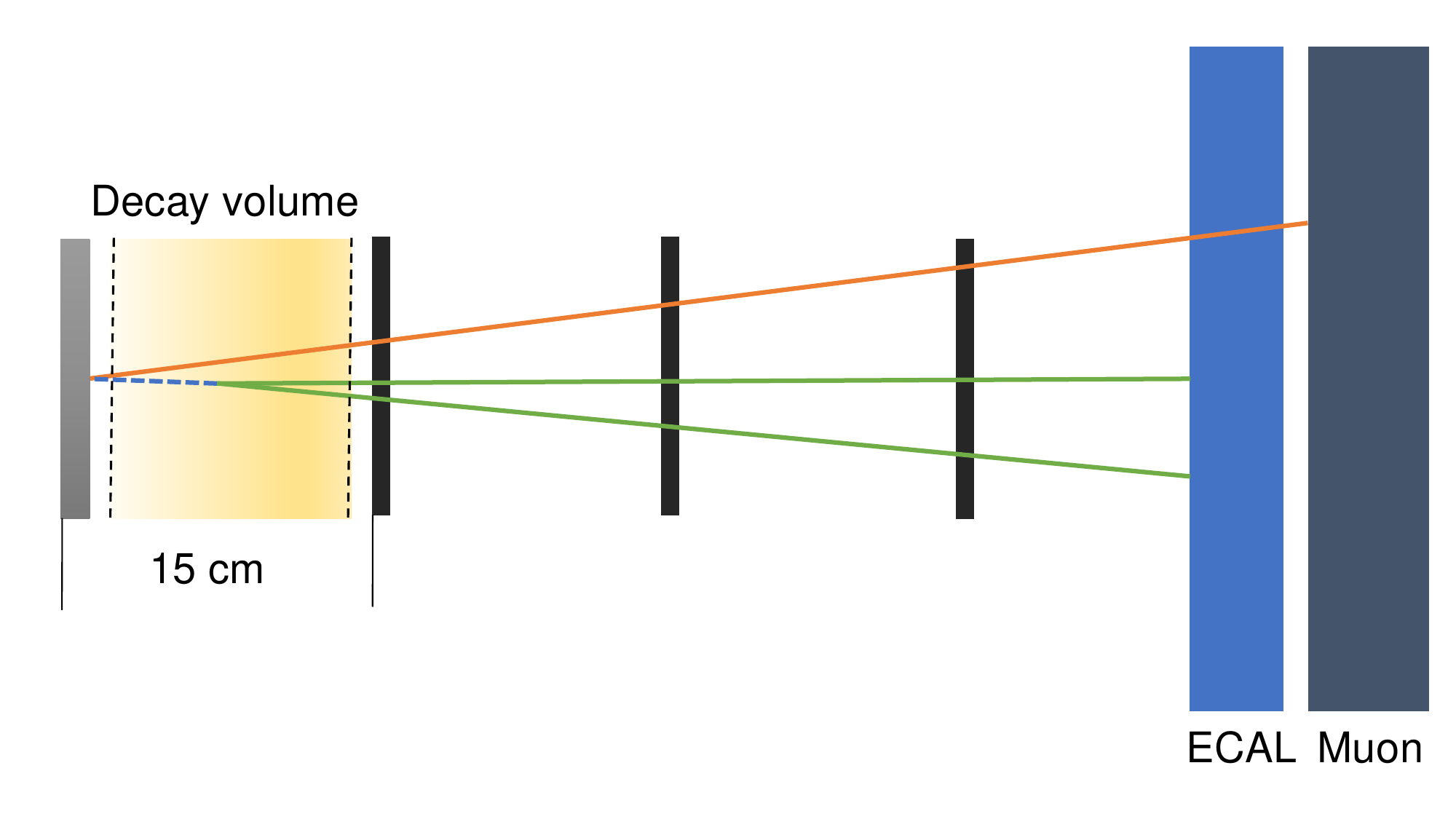}
    \vspace{-0.4cm}
    \caption{Signal event geometry in the corresponding volume, borrowed from Fig.~\cite{Krnjaic:2024ols}. We show the ECAL and muon filter after the module, but note that this can be a few meters away from the module that we consider.}
    \label{fig:signal}
\end{figure}

Fig.~\ref{fig:signal} shows the signal event process in a given module.
For a successful charged track reconstruction,
the tracks must pass through all three tracking stations in the module where they are produced.
This requires the displaced decay to happen before the first tracking station.
As shown in Ref.~\cite{Galon:2022xcl}, the fourth charged track from the primary electron does not need to be observed to distinguish signal events from backgrounds, so we inclusively search for the final state with or without this electron being captured.
For this reason, we do not show its track in Fig.~\ref{fig:signal}.
To avoid backgrounds from mis-reconstructing a prompt vertex as a displaced vertex, the displaced vertex should be at least $10 \delta z$ away from both the target and the first tracking station, with $\delta z \simeq 1~\rm mm$ being the reconstruction resolution along the $z$ direction (i.e. the streamline direction)~\cite{MUonE:2019qlm,Galon:2022xcl,Krnjaic:2024ols,Umberto:2021private}.
The decay volume is therefore defined as
\begin{align}
    25~\mathrm{mm} \leq z_{V\,\rm decay} \leq 140~\rm mm\,,
\end{align}
where $z_{V\,\rm decay}$ stands for the decay location of the vector boson along $z$ coordinate.
We further require the decay products to have an opening angle larger than 1 mrad to ensure accurate reconstruction.

\begin{table}[t!]
    \centering
    \begin{tabular}{|cc|}
        \hline
        Variable                     & Selection Criteria        \\
        \hline
        Decay $z$ coordinate         & $25$\,mm  $< z < 140$\,mm
        \\
        Decay daughter energy        & $> 1$\,GeV
        \\
        Decay daughter opening angle & $> 1$ mrad                \\
        Charged track geometry       & Hit all 3 trackers        \\
        Modules                      & Last 5                    \\
        \hline
    \end{tabular}
    \caption{Event selection criteria for our proposed search. In our numerical results, these requirements are imposed on our signal and background MC events. These criteria ensure that SM backgrounds are negligible for our signal of interest (see Supplemental Material for a discussion).}
    \label{tab:cuts}
\end{table}

For the purpose of background rejection, we require that all three charged tracks should enter the ECAL/muon filter for a successful PID.
Energy loss for charged particles penetrating materials may significantly impact the PID efficiency.
For example, an electron with energy 1 GeV will lose approximately 5\% of its energy when passing through one single Be target, see Ref.~\cite{walters2017stopping,ParticleDataGroup:2024cfk} for more details.
For this reason, we only use the last 5 modules of the streamline as a conservative choice.
Charged tracks that pass through all three tracking stations in these modules are guaranteed to be within the angular acceptance of the ECAL.
An energy threshold of 1 GeV is applied to the muon and the electron-positron pair from displaced decay as a mimic of the detector response to avoid potential reconstruction errors.

We summarize the proposed search strategy in Table~\ref{tab:cuts}.

\subsection{Signal Event Generation}
\label{sec:generation}

To evaluate the sensitivity of MUonE to hidden vector bosons, we generate 20000 $\mu^- N \to \mu^- N V$ and $\mu^- e^- \to \mu ^- e^- V$ events with a beam energy of 160\,GeV using \verb|MadGraph 3.2|~\cite{Alwall:2011uj,Alwall:2014hca,Frixione:2021zdp,Mattelaer:2021xdr}, with the model files constructed by \verb|FeynRules|~\cite{Alloul:2013bka}.
These events are weighted by the nuclear form factor, which dresses the electromagnetic vertex of the nucleus with $G(t)_2^{1/2} = \sqrt{G^{\rm el}_2 + G^{\rm in}_2}$, where
\begin{align}
    G^{\rm el}_2(t) & = \left( \frac{a^2t}{1 + a^2 t} \right)^2 Z^2 \left( \frac{1}{1+t/d} \right)^2\,,
    \\
    G_2^{\rm in}(t) & = \left(\frac{a'^2 t}{1+a'^2 t}\right)^2 Z \left(\frac{1 + t(\mu_p^2-1)/(4m_p^2)}{(1+t/(0.71~\rm GeV^2))^4}\right)^2\,.
\end{align}
For Beryllium, $a = 106Z^{-1/3}/m_e$, $d = 0.164$\,GeV$^2 A^{-2/3}$ for the elastic portion and $a' = 571 Z^{-2/3}/m_e$ and $\mu_p = 2.79$ for the inelastic portion \cite{Chen:2017awl, Tsai:1973py}.
This form factor is valid in the limit $t \ll m_p^2$~\cite{Bjorken:2009mm}.
The \verb|MadGraph| output contains very few events that violate this condition, but nonetheless, we discard those that approach it ($t > m_p^2/2$) by setting their weight to zero. We find that this has a negligible effect on the resulting bound.

\begin{figure}
    \centering
    \hspace{-1cm}
    \includegraphics[width=0.7\linewidth]{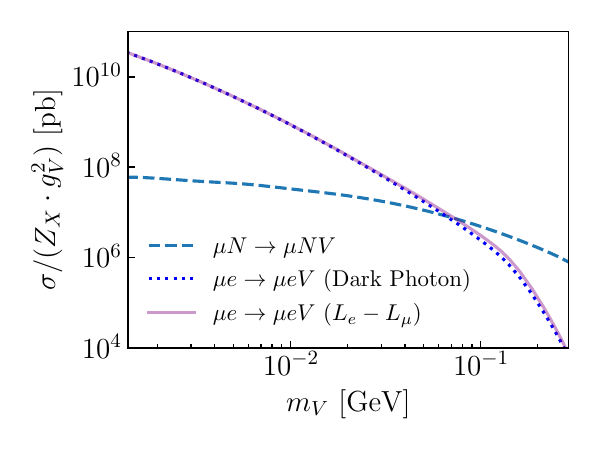}
    \vspace{-0.6cm}
    \caption{Production cross sections per unit charge, normalized to $g_V^2 = 1$ (Table.~\ref{tab:model}). At low masses, production from scattering off electrons is much more likely, as the electron is more likely to radiate out a gauge boson than the heavy nucleus, leading to a larger cross section. At large vector boson masses, production from nuclear scattering dominates, extending the experimental reach of MUonE to $\sim$ 100-150\, MeV.}
    \label{fig:productionCrossSections}
\end{figure}

Fig.~\ref{fig:productionCrossSections} shows the production cross section for the vector gauge boson in the $\mu e$ and $\mu N$ channel as a function of the mass.
The scaling behavior differs between the two production channels due to the inversion of the mass hierarchy between the target and beam.
As shown in the plot, $\mu e$ scattering dominates the cross section when the vector boson mass is below $\sim 60~\rm MeV$ because the target electron contributes significantly to the bremsstrahlung process, an effect which is heavily suppressed for the nuclear scattering case.
On the other hand, $\mu N$ scattering dominates in the large-mass regime because of the larger CoM energy for the nuclear target.
With a beam energy of 160\,GeV, the CoM energy in the nuclear scattering case is $\sqrt{s_{\mu N}} =54$\,GeV, compared to the electron scattering at just $\sqrt{s_{\mu e}} = 400$\,MeV.
The scattering off the nucleus thus enjoys a much wider available phase space, important for the production of a heavy vector boson.
An important consequence of this is that adding the $\mu N$ scattering significantly increases the maximum vector boson mass that MUonE is sensitive to.
In addition, as noted in Section~\ref{sec:model}, the relative sign of the electron and muon couplings to the dark vector boson in the $U(1)_{L_e - L_\mu}$ model can change the interference between diagrams in $\mu e \to \mu eV$.
Numerically, we find this effect to give an $\mathcal O(10) \%$ correction.

We comment on the analytical calculation of the cross section with Weizsaecker-Williams (WW) approximation~\cite{vonWeizsacker:1934nji,Williams:1934ad}.
Using the results of Ref.~\cite{Liu:2017htz} with the WW approximation, we find a factor of 1-2 difference in cross section with the results from \verb|MadGraph|. However, we find a larger discrepancy of a factor of $\sim 8$ between the \verb|MadGraph| result and the ``Improved WW'' technique used by Ref.~\cite{GrillidiCortona:2022kbq} to construct analytic MUonE bounds. This discrepancy could be explained by the assumption that $m_{beam} \ll m_{A'}$ that is necessary for the derivation of the cross section in Ref.~\cite{Liu:2017htz} (see e.g.~Ref~\cite{Bjorken:2009mm}), but invalid for the muon-flavored beam used by MUonE.
The difference in predicted cross sections partly explains the difference in projected sensitivities in this work and in Ref.~\cite{GrillidiCortona:2022kbq}, with the other differences discussed in Sec.~\ref{sec:results}.

\begin{figure}[t!]
    \centering
    \vspace{-0.5cm}
    \includegraphics[width=0.4\linewidth]{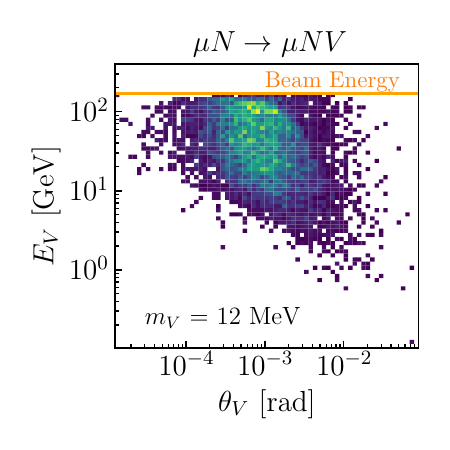} \hspace{-0.4cm} %
    \includegraphics[width=0.4\linewidth]{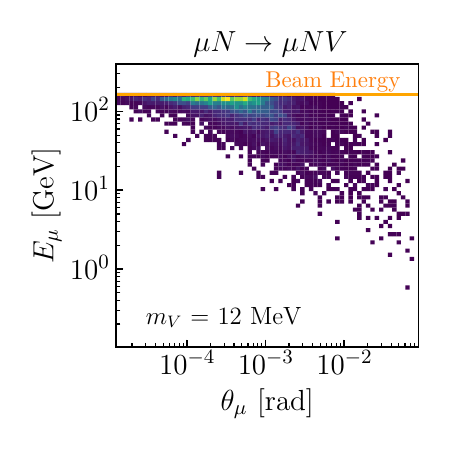} \hspace{-0.4cm}
    \vspace{-0.4cm}\\
    \includegraphics[width=0.4\linewidth]{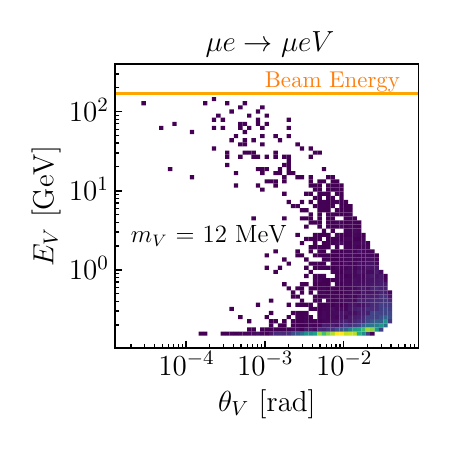} \hspace{-0.4cm} %
    \includegraphics[width=0.4\linewidth]{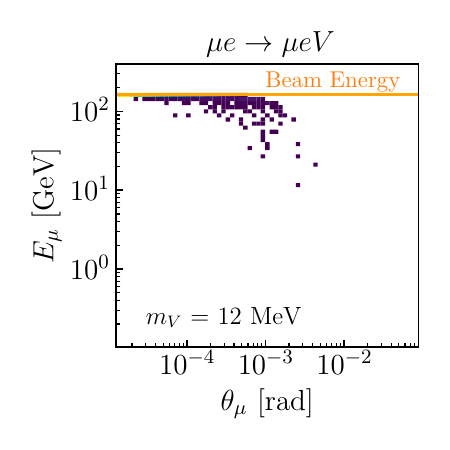} \hspace{-0.4cm}
    \vspace{-0.4cm}
    \caption{Energy and angle distribution for the produced vector boson and deflected muon for $m_{A'} = 12$ MeV.}
    \label{fig:kinem12}
\end{figure}

\begin{figure}[t!]
    \centering
    \vspace{-0.8cm}
    \includegraphics[width=0.4\linewidth]{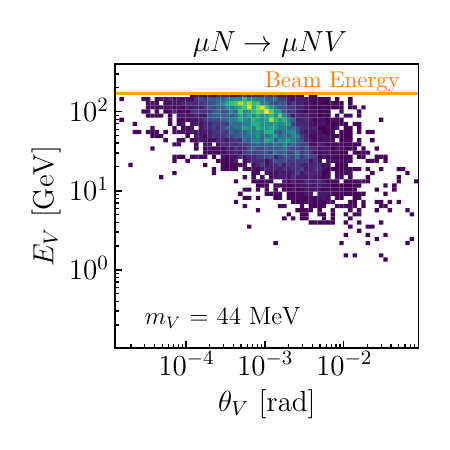} \hspace{-0.4cm} %
    \includegraphics[width=0.4\linewidth]{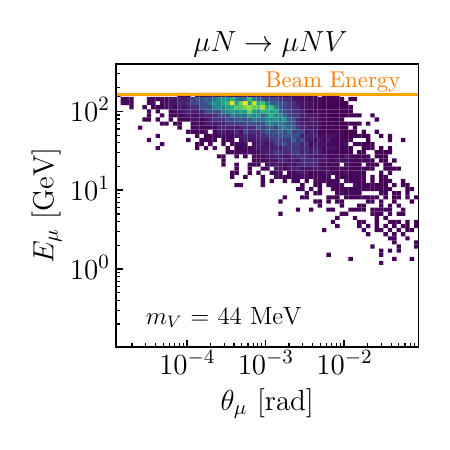} \hspace{-0.4cm}
    \vspace{-0.4cm}\\
    \includegraphics[width=0.4\linewidth]{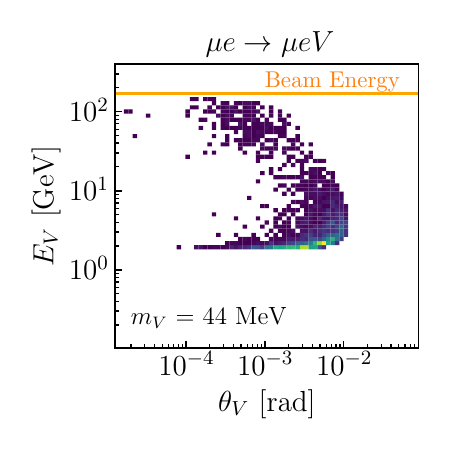} \hspace{-0.4cm} %
    \includegraphics[width=0.4\linewidth]{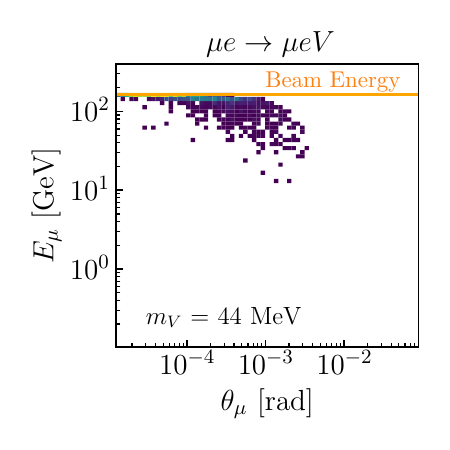} \hspace{-0.4cm}
    \vspace{-0.4cm}
    \caption{Energy and angle distribution for the produced vector boson and deflected muon for $m_{A'} = 44$ MeV.}
    \label{fig:kinem83}
\end{figure}

Figs.~\ref{fig:kinem12}~and~\ref{fig:kinem83} display the
kinematic distributions of the products in $\mu e \to \mu e V$ and $\mu N \to \mu NV$ scatterings for $m_V = 12~\rm MeV$ and $m_V = 44~\rm MeV$.
As seen in the figures, scattering off electrons produces much softer vector gauge bosons than those produced from scattering off nuclei.
The energy cut of $E_{\rm prod} > 1$\,GeV implies that vector bosons of $E_V < 2$\,GeV are completely invisible. Above this threshold, the probability of passing the $E_{\rm prod}$ cut for both resulting electrons increases and smoothly approaches unity at high $E_V$.
For $m_V = 12~\rm MeV$, a significant portion of the produced vector bosons lie well below 2\,GeV, leading to a low detection efficiency.
This offsets the much greater production cross section of the electron channel, so both channels yield a similar number of signal events, as shown later in Sec.~\ref{sec:results}.
For $m_V = 44~\rm MeV$, the energies of the vector bosons produced in $\mu e$ scattering are above 2 GeV, which improves the efficiency.
However, at $m_V = 44$\,MeV, the production cross sections for both channels are comparable, again resulting in a similar contribution to signal events from the nuclear and electron scattering channels.

After production via dark bremsstrahlung, the vector boson decays, with decay length given in Eq.~\ref{eq:decay fermion}.
The kinematics of the decay are computed manually by boosting to the rest frame of the vector boson, where the decay occurs isotropically, and then inverse boosting.
In the regions of parameter space excluded by MUonE, the $m_V < 2 m_\mu$, so the only available decay modes at tree-level are to electrons or neutrinos.
Electrons are the only visible channel, and the visible branching ratios (for $m_A \gg 2m_e$) are $1, 2/5, 1/2$ for the dark photon, $U(1)_{B-L}$, and $U(1)_{L_e - L_\mu}$ models, respectively.
The lower branching ratio to visible products can slightly reduce the reach of MUonE, see Sec.~\ref{sec:results} for a discussion.
Below $m_{A'}<2m_e$, the dark vector decays to neutrinos or is very nearly stable, and both scenarios are invisible to MUonE. After decay, the resulting visible leptons are required to pass the cuts in Table.~\ref{tab:cuts}, and we compile those that pass into an expected number of signal events.

\subsection{Backgrounds}
\label{sec:background}

The background falls into two categories: (a) SM processes that happen within the target or the first tracking station and create a prompt vertex, which may be mis-reconstructed as a displaced vertex due to limited resolution. (b) SM processes that can naturally produce a displaced vertex.
The first category can be safely rejected by requiring a $10\delta z$ displacement away from the target and the first tracking station, while the second category requires a simulation to examine its impact.

Ref.~\cite{Galon:2022xcl} performed a systematic, simulation-based analysis for the background process with an inclusive fourth track.
It was concluded that the backgrounds can all be safely rejected.
Most hadronic processes come with many extra charged particles that leave hits on the tracking stations or the ECAL, leading to rejection.
The remaining small portion of events is relatively clean, and the PID system is required to reject them.
The vast majority of background events arise from $K^0 \to \pi^+ \pi^-$ and $\Lambda \to p \pi^-$.
The simulation finds that about 4000 such background events are expected to fake the signal topology, and can be safely rejected by ECAL as long as the per-particle-fake rate is less than 1.6\%.
A small number of events have $K^0 \to \pi e \nu$ or $\Lambda \to p e \nu$, where one real electron/positron is faking the displaced decay and the other is from a charged pion or proton.
Simulation finds that about 8 events are expected, and can be rejected by the ECAL.
Finally, about 0.5 events are expected to have two real electrons faking the displaced decay, such as $K^0 \to \pi^0 \pi^0$ with one $\pi^0$ decaying into $e^+ e^- \gamma$.
Fortunately, all these events have at least one non-electron particle hitting the ECAL (usually a charged pion from the primary vertex, or an energetic photon), leading these events to be rejected.
As a conclusion, all backgrounds can be safely rejected by our search strategy, and we assume the backgrounds are zero in our statistics.

\section{Analysis and Results}
\label{sec:results}

Based on the discussion of Sec.~\ref{sec:signal}, we perform a parameter scan over the coupling-mass plane.
We generate signal event samples for 16 mass points logarithmically spaced between 2 MeV and 300 MeV.
For each mass point, we apply the selection criteria on an event-by-event basis for couplings ranging from $10^{-6}$--$1$, covering 30 logarithmically spaced points.
For each parameter point, the total number of signal events is calculated by
\begin{align}
    N_{\rm sig} = \epsilon_{\mathrm{sig}, N} \sigma_N \mathcal{L}_N + \epsilon_{\mathrm{sig}, e} \sigma_e \mathcal{L}_e\,.
\end{align}
Here $\epsilon_{\mathrm{sig}, N(e)}$ is the fraction of signal events from $\mu N$ ($\mu e$) scattering that pass the cuts in Table.~\ref{tab:cuts}, $\sigma_{N, e}$ is the production cross section calculated using \verb|MadGraph|, and $\mathcal{L}_{N,e}$ is the effective luminosity.
For $\mu N$ scattering, $\mathcal{L}_N = 1.5 \times 10^{4} /  8 /4 ~\rm pb^{-1}$, while for $\mu e$ scattering we have $\mathcal{L}_e = 1.5 \times 10^{4} /  8~\rm pb^{-1}$.
The factor of 8 comes from the fact that only last 5 modules can be used (Sec.~\ref{sec:search}), while the factor of 4 in $\mathcal{L}_N$ comes from the Be atomic number.

\begin{figure}[t!]
    \centering
    \includegraphics[width=\linewidth]{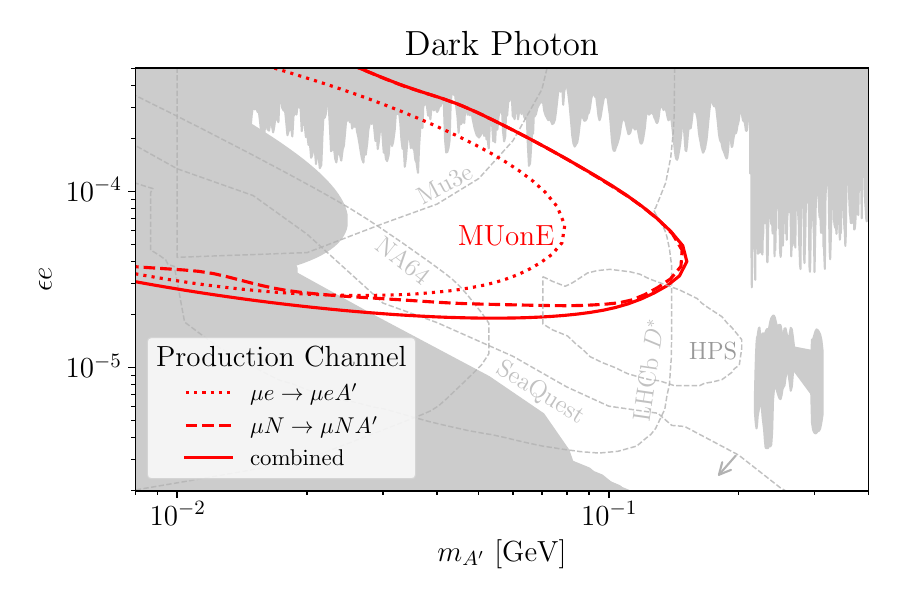}
    \caption{The result for dark photon search. MUonE sensitivity is shown with the red dotted curve (for $\mu e$ scattering), the red dashed curve (for $\mu N$ scattering), and the red solid curve (combined). Existing limits are shown in gray shaded regions. Selected future limits that are planned for a similar timeline as MUonE are also shown. Specifically, we add an arrow for the SeaQuest projection to indicate that it is expected to probe the parameter space below the curve.}
    \label{fig:dark photon result}
\end{figure}

\begin{figure}[t!]
    \centering
    \includegraphics[width=\linewidth]{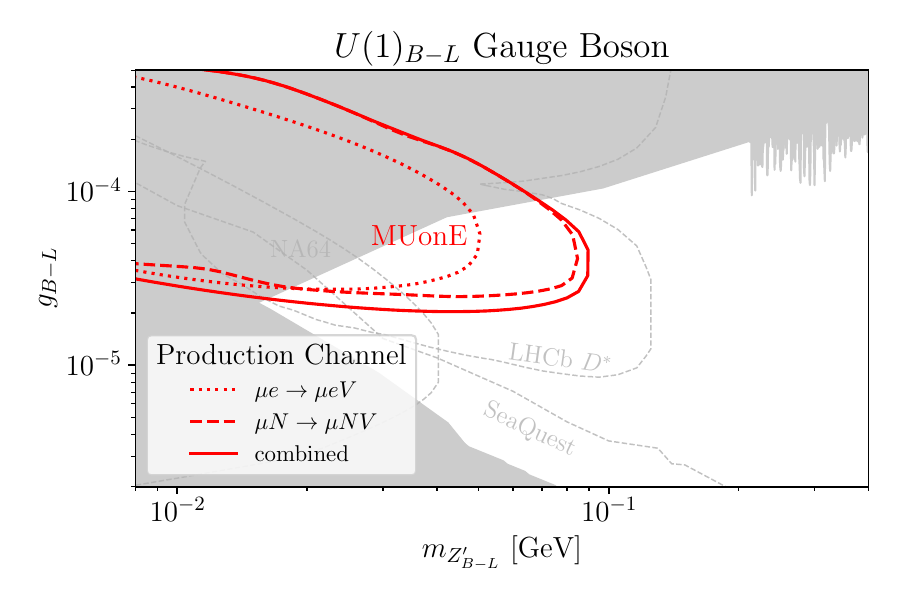}
    \vspace*{-1.0cm}
    \caption{The result for gauged $U(1)_{B-L}$ search. MUonE sensitivity is shown with the red dotted curve (for $\mu e$ scattering), the red dashed curve (for $\mu N$ scattering), and the red solid curve (combined). Existing limits are shown in gray shaded regions. Selected future limits that are planned for a similar timeline as MUonE are also shown. Specifically, we add an arrow for the SeaQuest projection to indicate that it is expected to probe the parameter space below the curve.}
    \vspace*{-0.5cm}
    \label{fig:B-L result}
\end{figure}

\cref{fig:dark photon result,fig:B-L result,fig:Le-Lmu result} show our main results for the dark photon, $U(1)_{B-L}$ gauge boson, and $U(1)_{L_e - L_\mu}$ gauge boson.
For each plot, we show the projected sensitivity at 95\% confidence level assuming zero background from $\mu e$ and $\mu N$ scattering with red dotted and dashed curves, respectively.
The combined result is shown with the red solid curve.
In all cases, we find that the $\mu e$ scattering can reach a lower coupling than the $\mu N$ scattering with vector boson mass below $\sim 30~\rm MeV$.
This is due to the larger energy acquired by the vector boson from the Be nucleus as seen in~\cref{fig:kinem12,fig:kinem83}, which leads to a longer decay length that should be compensated by a larger coupling.
The $\mu N$ scattering, on the other hand, can reach a much higher vector boson mass because of the larger CoM energy.

The existing limits are shown in the gray shaded area, while future sensitivities within the next 5 years are shown by gray dashed curves.
Our recasting of these limits relies on \textsc{darkcast}~\cite{Ilten:2018crw}.
We also refer the readers to Ref.~\cite{Bauer:2018onh} for a comprehensive study on $U(1)$ vector bosons.
Note that all of our searches are based on the assumption that the vector bosons only decay back to SM particles.
Invisible searches, assuming decays into stable dark sector particles, do not apply.

In the following subsections, we comment on our results and emphasize the differences among these models.

\subsection{Dark photon model}

Fig.~\ref{fig:dark photon result} shows the MUonE projection for dark photon searches.
We found that the MUonE reach extends to dark photon masses as heavy as $\sim 150~\rm MeV$ for couplings that fall into the ``gap''.
Note that the sensitivity is significantly smaller than that in Ref.~\cite{GrillidiCortona:2022kbq}.
Besides the discussion of the WW approximation in Sec.~\ref{sec:generation},
the restriction to only the last 5 modules suppresses the luminosity.
We point out that this is essential to have a reliable PID accuracy for background rejection.
Existing limits are from $(g-2)_e$~\cite{Pospelov:2008zw,Endo:2012hp}, beam dumps~\cite{Riordan:1987aw,Bjorken:1988as,Bross:1989mp,Konaka:1986cb,Davier:1989wz,Bjorken:2009mm,Andreas:2012mt,Blumlein:1990ay,Blumlein:1991xh,Blumlein:2011mv,Blumlein:2013cua,Bergsma:1985qz,Gninenko:2012eq,Astier:2001ck,Bernardi:1985ny,Gninenko:2011uv,NA64:2018lsq,NA64:2019auh}), electron fixed-target~\cite{Merkel:2014avp,Abrahamyan:2011gv}, collider experiments~\cite{Aubert:2009cp,Curtin:2013fra,Lees:2014xha,Ablikim:2017aab,Aaij:2017rft,Archilli:2011zc,KLOE:2016lwm,FASER:2023tle}, and rare meson decays~\cite{Bernardi:1985ny,MeijerDrees:1992kd,Archilli:2011zc,Gninenko:2011uv,Babusci:2012cr,Adlarson:2013eza,Agakishiev:2013fwl,Adare:2014mgk,Batley:2015lha,KLOE:2016lwm}.
Future projections shown here include HPS~\cite{HPS:2018xkw,Baltzell:2022rpd}, NA64$e$ with $10^{12}$ electrons on target~\cite{Gninenko:2300189}, $D^*$ decay search at LHCb upgrade~\cite{Ilten:2015hya,Ilten:2016tkc}, SeaQuest~\cite{Gardner:2015wea} ,and Mu3e~\cite{Echenard:2014lma}.

\newpage
\subsection{The gauged $U(1)_{B-L}$ model}

Fig.~\ref{fig:B-L result} shows the MUonE sensitivity to the $U(1)_{B-L}$ gauge boson.
The existing dark photon limits and projections discussed in the previous subsection are re-interpreted for the $U(1)_{B-L}$ gauge boson using \textsc{darkcast}~\cite{Ilten:2018crw}.

The main difference arises from the coupling to the neutrinos.
This opens an additional decay channel into a neutrino pair, shortening the decay length by a factor of $2/5$ compared to the dark photon model.
To maintain the same decay length, the coupling must be reduced slightly, which in turn lowers the cross section.
Moreover, since neutrinos are invisible, the observable signal rate is further suppressed by the branching ratio into $e^+ e^-$, which carries the same factor of $2/5$.
As a result, the largest coupling that is accessible for MUonE for a given mass drops by an $\mathcal{O}(1)$ factor, and the maximum mass also reduces to about $ 100~\rm MeV$, slightly weaker than in the dark photon case.
In addition, the $U(1)_{B-L}$ gauge theory is further constrained by neutrino experiments such as Borexino~\cite{Bellini:2011rx,Harnik:2012ni}, Texono~\cite{Bilmis:2015lja}, CHARM II~\cite{CHARM-II:1993phx}, SPEAR, DORIS, and PETRA~\cite{Carlson:1986cu,Frugiuele:2016rii}, resulting in the extra gray-shaded region for masses around a few tens of MeV.
These additional limits cover the projection of Mu3e, so it is not shown in Fig.~\ref{fig:B-L result}.
We note that the HPS projection is absent due to the lack of sufficient information to translate the result into this model.

\subsection{The gauged $U(1)_{L_e - L_\mu}$ model}
\begin{figure}[t!]
    \centering
    \includegraphics[width=\linewidth]{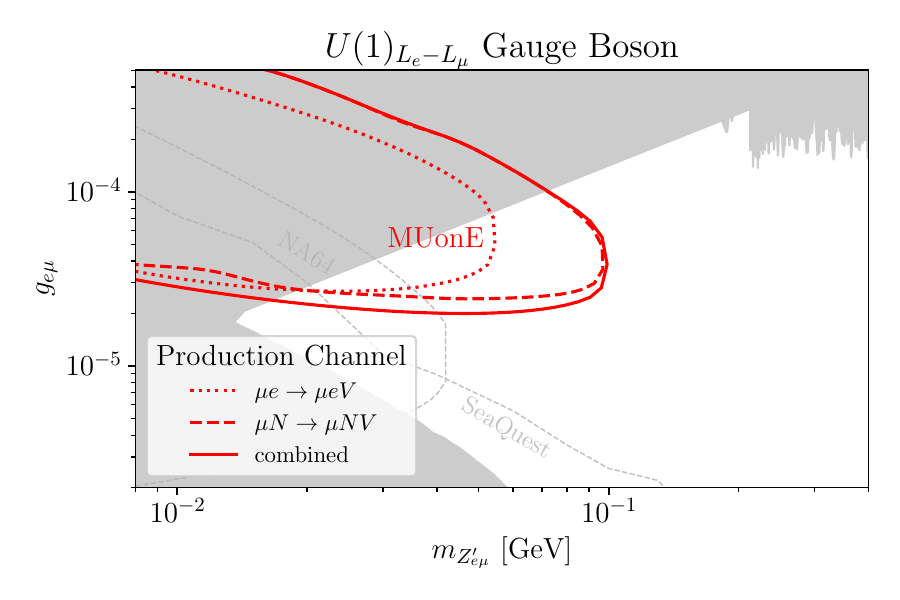}
    \vspace*{-1.0cm}
    \caption{The result for gauged $U(1)_{L_e - L_\mu}$ search. MUonE sensitivity is shown with the red dotted curve (for $\mu e$ scattering), the red dashed curve (for $\mu N$ scattering), and the red solid curve (combined). Existing limits are shown in gray shaded regions. Selected future limits that are planned for a similar timeline as MUonE are also shown. Specifically, we add an arrow for the SeaQuest projection to indicate that it is expected to probe the parameter space below the curve.}
    \label{fig:Le-Lmu result}
\end{figure}
Fig.~\ref{fig:Le-Lmu result} shows the MUonE sensitivity for the gauged $U(1)_{L_\mu - L_e}$ model, with existing dark photon limits re-interpreted by running \textsc{darkcast}~\cite{Ilten:2018crw}.
The discussion on the impact of neutrino coupling for the $U(1)_{B-L}$ model also applies here, though this gauge boson couples to one fewer neutrino flavor.
This slightly enhances the sensitivity over the $U(1)_{B-L}$ case.

Unlike the previous two models, the $U(1)_{L_e - L_\mu}$ gauge boson does not couple to quarks at tree-level (although it can do so at loop-level).
Consequently, the production of $U(1)_{L_\mu - L_e}$ gauge boson in hadronic experiments such as CHARM, NuCAL, and NOMAD is strongly suppressed.
Nevertheless, electron beam-dump experiments have achieved similar sensitivities, so the existing bounds are not strongly affected.

The absence of a $\tau$ neutrino coupling suppresses the sensitivity of many neutrino experiments, but raises it in Super-K due to modified neutrino oscillations~\cite{Wise:2018rnb}.
Among the neutrino experiments mentioned in the $U(1)_{B-L}$ case, the Texono~\cite{Bilmis:2015lja} and Borexino~\cite{Bellini:2011rx,Harnik:2012ni} remain relevant as their primary search channel is through scattering between electrons and electron neutrinos.
As the $U(1)_{B-L}$ case, the Mu3e and HPS projections are absent for the reasons explained above.

\section{Conclusion}
\label{sec:conclusion}

In this paper, we demonstrate that the MUonE experiment provides a unique opportunity to probe hidden vector bosons through displaced decay signatures.
In exploring the dark photon, $U(1)_{B-L}$, and $U(1)_{L_e - L_\mu}$ models as benchmark examples, we extended the prior work of Ref.~\cite{Galon:2022xcl} to include the nuclear scattering, $\mu N \to \mu NV$, in addition to the $\mu e \to \mu e V$ channel.
We perform a dedicated simulation of signal events and apply the event selection criteria.
Our analysis illustrates that $\mu N$ scattering, owing to its higher CoM energy, substantially enhances the reach of MUonE for heavier mediators, extending the sensitivity up to $m_V \sim 150$\,MeV, a factor of two improvement over the case with $\mu e$ scattering alone.
Together with the performance of $\mu e$ scattering at low mass, the combined production channels show a promising sensitivity.

Of course, more BSM scenarios could potentially be probed at MUonE, and this work serves as a proof of concept.
As MUonE approaches its data-taking phase, we hope this work provides a more solid assessment of its discovery potential, and thus motivates future explorations in the experiment.

\section*{Acknowledgement}
We are grateful to Gordan Krnjaic for helpful discussions, feedback on the manuscript, and for encouraging this paper to be written.
We thank Tao Han for useful comments.
Fermilab is operated by Fermi Forward Discovery Group, LLC under Contract No. 89243024CSC000002 with the U.S. Department of Energy, Office of Science, Office of High Energy Physics.
I.R.W. is also supported by DOE distinguished scientist fellowship grant FNAL 22-33.

\bibliographystyle{JHEP}
\bibliography{main}
\end{document}